\documentclass[aps,prd,twocolumn,superscriptaddress,showkeys,nofootinbib,notitlepage,amsmath,amssymb,preprintnumbers]{revtex4-2}

\usepackage[english]{babel}
\usepackage{xcolor}
\usepackage{mathtools}
\usepackage{pdfsync}
\usepackage{amssymb}
\usepackage{amsmath}
\usepackage{hyperref}
\usepackage{bm}
\usepackage{mathtools}
\usepackage{slashed}
\usepackage[active]{srcltx}
\def\d{\hbox{{d}\kern-.20em\hbox{l}}}

\def \matrix #1 {\left(\begin{array}{cc} #1 \end{array}\right)}

\def\II{\hbox{{1}\kern-.25em\hbox{l}}}

\newcommand \widebar [1] {\overline{#1}}

\begin{document}


\title{
On evolution kernels of twist-two operators
}

\author{Yao Ji}
   \email{yao.ji@tum.de}
   \affiliation{Physik  Department  T31,  
   Technische  Universit\"at  M\"unchen,  D-85748  Garching, Germany}

\author{Alexander  Manashov}
   \email{manashov@mpp.mpg.de}
\affiliation{Max-Planck-Institut f\"ur Physik, Werner-Heisenberg-Institut, D-80805 M\"unchen, Germany}

\author{Sven-Olaf Moch}
   \email{sven-olaf.moch@desy.de}
\affiliation{II. Institut f\"ur Theoretische Physik, Universit\"at Hamburg
   D-22761 Hamburg, Germany}


\begin{abstract}

The evolution kernels that govern the scale dependence of the generalized parton distributions are invariant under transformations of the
$\mathrm{SL}(2,\mathrm R)$ collinear subgroup of the conformal group. Beyond one loop the symmetry generators, due to quantum effects,
differ from the canonical ones. We construct the transformation which brings the {\it full} symmetry generators back to their canonical
form and show that the eigenvalues (anomalous dimensions) of the new, canonically invariant, evolution kernel coincide with the so-called
parity respecting anomalous dimensions. We develop an efficient method that allows one to restore an invariant kernel from the
corresponding anomalous dimensions. As an example, the explicit expressions for NNLO invariant kernels for the twist two
flavor-nonsinglet operators in QCD and for the planar part of the universal anomalous dimension in ${\cal N}=4$ SYM are presented.

\end{abstract}

\preprint{ \textmd{\small TUM--HEP--1461/23,\
 MPP--2023--137, \ DESY--23--091}}

\keywords{
evolution kernels, DVCS, conformal symmetry, generalized parton distribution}

\maketitle

%
%
\section{Introduction}\label{sect:Intro}
The study of deeply-virtual Compton scattering (DVCS) gives one  access  to  the  generalized  parton
distributions~\cite{Muller:1994ses,Radyushkin:1996nd,Ji:1996nm} (GPDs)  that  encode  the  information on the transverse position of quarks
and gluons in the proton in dependence on their longitudinal momentum. In order to extract the GPDs from  experimental data one has to
know, among other things, their scale dependence. The latter is governed by the renormalization group equations (RGEs) or, equivalently,
evolution equations for the corresponding twist two operators. Essentially the same equations govern the scale dependence of the ordinary
parton distribution functions (PDFs) in the Deep Inelastic Scattering (DIS) process. In DIS one is interested in the scale dependence of
forward matrix elements of the local twist-2 operators and therefore can neglect the operator mixing problem between local operators from
the operator product expansion (OPE).  In the nonsinglet sector, there is only one operator for a given spin/dimension. The anomalous
dimensions of such operators are known currently  with the three-loop accuracy~\cite{Moch:2004pa,Vogt:2004mw} and first results at four
loops are becoming available~\cite{Moch:2017uml,Falcioni:2023luc}. In contrast, the DVCS process corresponds to non-zero momentum transfer
from the initial to the final state and, as a consequence, the total derivatives of the local twist-two operators have to be taken into
consideration. All these operators mix under renormalization and the RGE has a matrix form. The DIS anomalous dimensions appear as the
diagonal entries of the anomalous dimensions matrix which, in general, has a triangular form for the latter.

It was shown by Dieter M\"uller~\cite{Mueller:1991gd,Mueller:1997ak} that the off-diagonal part of the anomalous dimension matrix is
completely determined by a special object, the so-called  conformal anomaly. Moreover, in order to determine the off-diagonal part of
the anomalous dimension matrix with  $\ell$-loop accuracy it is enough to calculate the conformal anomaly at one loop less. This technique was
used to reconstruct all relevant evolution kernels/anomalous dimension matrices in QCD at two
loops~\cite{Belitsky:1997rh,Belitsky:1998gc,Belitsky:1999hf}.

A similar approach, but based on the analysis of QCD at the critical point in non-integer dimensions, was developed in
refs.~\cite{Braun:2013tva,Braun:2014vba,Braun:2016qlg}. It was shown that the evolution kernels in $d=4$ in the
$\overline{\mathrm{MS}}$-like renormalization scheme inherit the symmetries of the critical theory in $d=4-2\epsilon$ dimensions.  As
expected, the symmetry generators deviate from their canonical form. Corrections to the generators have a rather simple form if they are
written in terms  of the evolution kernel  and the conformal anomaly. It was shown in ref.~\cite{Braun:2017cih} that by changing a
renormalization scheme one can get rid of the conformal anomaly term in the generators bringing them into the so-called ``minimal" form.
Beyond computing the evolution kernels, the conformal approach has also been employed to calculate the NNLO coefficient (hard) functions of
vector and axial-vector contributions in DVCS~\cite{Braun:2020yib,Braun:2021grd}, the latter in agreement with a direct Feynman diagram
calculation~\cite{Gao:2021iqq}. Moreover, the conformal  technique is also applicable to computing kinematic higher-power corrections in
two-photon processes as was recently shown in refs.~\cite{Braun:2020zjm,Braun:2022qly}.

In this paper we construct a similarity transformation that  brings the full quantum generators back to the canonical form.
Correspondingly, the transformed evolution kernel is invariant under the canonical $\mathrm{SL}(2,\mathrm R)$ transformation. Moreover, we
will show that the eigenvalues of this kernel are given by the so-called parity respecting anomalous dimension,
$f(N)$~\cite{Basso:2006nk,Dokshitzer:2005bf} which is related to the PDF anomalous dimension spectrum $\gamma(N)$ as
\begin{align}
\label{eq:no1}
\gamma(N)=f\left(N+{\bar\beta}(a) +\frac12 \gamma(N)\right),
\end{align}
where $\bar\beta(a)=-\beta(a)/2a$ with $\beta(a)$ being the QCD  beta function. The strong coupling $\alpha_s$ is normalized as
$a=\alpha_s/(4\pi)$. We develop an effective approach to restore the canonically invariant kernel from its eigenvalues $\gamma(N)$. As an
example, we present explicit expressions for three-loop invariant kernels in QCD and ${\cal N}=4$ supersymmetric Yang-Mills (SYM) theory.
The answers are given by linear combinations of harmonic polylogarithms~\cite{Remiddi:1999ew}, up to weight four in QCD and up to weight
three in ${\cal N}=4$ SYM. We also compare our exact result with the approximate expression for the three-loop kernels in QCD given in
ref.~\cite{Braun:2017cih}.

The paper is organized as follows: in section~\ref{sect:prelim} we describe the general structure of the evolution kernels of twist-two
operators. In section~\ref{sect:gammas} we explain how to effectively recover the evolution kernel from the known anomalous dimensions and
present our results for the invariant kernels in QCD and ${\cal N}=4$ SYM. Sect.~\ref{sect:summary} contains the concluding remarks. Some
technical details are given in the Appendices.

\section{Kernels \& symmetries}\label{sect:prelim}
We are interested in the scale dependence of the twist-two light-ray flavor nonsinglet operator~\cite{Balitsky:1987bk}
\begin{align}  \label{lightray}
\mathcal O(z_1,z_2) & = [\bar q(z_1n)\gamma_+ [z_1n,z_2n] q(z_2n)]_{\overline{\text{MS}}},
\end{align}
where $n^\mu$ is an auxiliary light-like vector, $n^2 = 0$, $z_{1,2}$ are real numbers, $\gamma_+  = n^\mu\gamma_\mu$ and  $[z_1n,z_2n]$
stands for the Wilson line ensuring gauge invariance, and the subscript $\widebar{\rm MS}$ denotes the renormalization 
scheme. This operator
 can be viewed as the generating function for local operators, ${\mathcal O}^{\mu_1\dots\mu_N}$ that are symmetric
and traceless in all Lorentz indices $\mu_1\dots\mu_N$.

The renormalized  light-ray operator~\eqref{lightray} satisfies the RGE
\begin{align}\label{VARGE}
\Big(\mu\partial_\mu +\beta(a)\partial_{a}+ \mathbb{H}(a)\Big) {\mathcal O}(z_1,z_2)=0,
\end{align}
where 
$\beta(a)$ is $d$-dimensional beta function
\begin{align}
\beta(a)= -2a\big(\epsilon+\beta_0 a +\beta_1 a^2+O(a^3)\big),
\end{align}
$\beta_0=11/3 N_c-2/3n_f$, etc., and $\mathbb{H}(a)=a \mathbb H_1 + a^2 \mathbb H_2+\ldots$ is an integral operators in $z_1,z_2$.

It follows from the invariance of the classical QCD Lagrangian under conformal transformations that the one-loop kernel $\mathbb H_1$
commutes with the \textit{canonical} generators of the collinear conformal subgroup, $S_0, S_\pm$,
\begin{align}\label{Streelevel}
S_- & =-\partial_{z_1}-\partial_{z_2}\,,
\notag\\
S_0 & =z_1\partial_{z_1} +  z_2\partial_{z_2} + 2\,,
\notag\\
S_+ & =z_1^2\partial_{z_1} +  z_2^2\partial_{z_2} +2z_1 + 2z_2\,.
\end{align}
This symmetry is preserved beyond one loop albeit two of the  generators, $S_0, S_+$ receive quantum corrections, $S_\alpha\mapsto
\widetilde S_\alpha(a)=S_\alpha+\Delta S_\alpha(a)$. The explicit form of these
corrections can be found in ref.~\cite{Braun:2016qlg}.

It is quite useful to bring the generators to the following form using the similarity transformation~\cite{Braun:2017cih},
\begin{align}
\mathbb H(a)=e^{-X(a)} \mathrm H(a) e^{X(a)}\,, &&
\notag \\
\widetilde S_\alpha(a) =e^{-X(a)} \mathrm S_\alpha(a) e^{X(a)}\,,
\end{align}
where  $X(a)=a X_1 +a^2 X_2 +\ldots$ is an integral operator known up to terms of $O(a^3)$~\cite{Belitsky:1998gc,Braun:2017cih}. This
transformation can be thought of as a change in a renormalization scheme.

The shift operator $\mathrm S_-$ is not modified and hence identical to $S_-$ in Eq.~\eqref{Streelevel},  and the quantum corrections to
$\rm S_0$ and $\rm S_+$ come only through the evolution kernel
\begin{subequations}\label{Smathrm}
\begin{align}
\mathrm S_0(a) & =S_0+\bar\beta(a) +\frac12 \mathrm H(a)\,,\label{S0}
\\
\mathrm S_+(a) & =S_+ + (z_1 + z_2)\left(\bar\beta(a) +\frac12 \mathrm H(a)\right)\,,
\label{Splus}
\end{align}
\end{subequations}
where $\bar\beta(a)=\beta_0 a+\beta_1 a^2+\cdots$ is the beta function in four dimensions, cf. Eq.~\eqref{eq:no1}.
The form of the generator $\mathrm S_0(a)$ is completely fixed by the scale invariance
of the theory, while Eq.~\eqref{Splus} is the ``minimal" ansatz consistent with the commutation relation $[\mathrm S_+,\mathrm S_-]=2\mathrm
S_0$.  Since the operator $\mathrm H(a)$  commutes with the generators, $[\mathrm H(a),\mathrm S_\alpha(a)]=0$ its form is completely
determined by its spectrum (anomalous dimensions). However, since the
generators do not have the simple form as in Eq.~\eqref{Streelevel}, it is yet necessary to find
a way to recover the operator  from its spectrum.

To this end we construct a transformation which brings the generators $\mathrm S_\alpha(a)$ to the canonical form~$S_\alpha$,
Eq.~\eqref{Streelevel}. Let us define an operator $\mathrm T(\mathrm H)$:
\begin{align}\label{T(H)}
\mathrm T(\mathrm H)=\sum_{n=0}^\infty \frac1{n!}\mathrm L^n \left(\bar\beta(a)+\frac12 \mathrm H(a) \right)^n\,,
\end{align}
where $\mathrm L= \ln z_{12}$, $z_{12}\equiv z_1-z_2$. Recall that $z_1,z_2$ are real variables, so for $z_{12}<0$ it is necessary to
choose a specific branch of the logarithm function. Although this choice is irrelevant for further analysis we chose the $+i0$ recipe for
concreteness, i.e., $\mathrm L=\ln (z_{12}+i0)$. It can be shown that the operator $\mathrm T(\mathrm H)$ intertwines the 
symmetry generators $\mathrm S_\alpha(a)$ and the canonical generators,~$S_\alpha$. Namely,
\begin{align} \label{TintS}
\mathrm T(\mathrm H)\, \mathrm S_\alpha(a) = S_\alpha \,\mathrm T(\mathrm H),
\end{align}
see Appendix~\ref{appendix:A} for details. Let us also define a new kernel $\widehat{\mathrm H}$ as
\begin{align} \label{TintH}
\mathrm T(\mathrm H)\, \mathrm H(a) = \widehat{\mathrm H}(a) \,
     \mathrm T(\mathrm H).
\end{align}
It follows from Eqs.~\eqref{TintS}, \eqref{TintH} that the operator
$\widehat{\mathrm H}$ commutes with the canonical generators in
Eq.~\eqref{Streelevel}
\begin{align}\label{ScanH}
[S_\alpha,\widehat{\mathrm H}(a)]=0.
\end{align}
The problem of restoring a canonically invariant operator $\widehat{\rm H}(a)$ from its spectrum is much easier than that for the operator
$\mathrm H(a)$ and will be discussed in the next section.
%
It can be shown that the inverse of $\mathrm T(\mathrm H)$ takes the form
\begin{align}\label{inverseT}
\mathrm T^{-1}(\mathrm H) =
\sum_{n=0}^\infty \frac{(-1)^n}{n!} \mathrm L^n\left(\bar\beta(a)+\frac12\widehat{  \mathrm H}(a) \right)^n\,,
\end{align}
see Appendix~\ref{appendix:A}. Further, it follows from Eq.~\eqref{TintH} that
\begin{align}
    \label{HhatH}
\mathrm H(a) &=\mathrm T^{-1}(\mathrm H)\, \widehat{\mathrm H}(a)\,  \mathrm T(\mathrm H)
     \notag\\
     &=\widehat{\mathrm H}(a)
     +\sum_{n=1}^\infty \frac1{n!}\mathrm T_n(a)\left(\bar\beta(a)+\frac12{  \mathrm H}(a) \right)^n\,.
\end{align}
The operators $\mathrm T_{n}(a)$ are defined by recursion
\begin{align}
\mathrm T_n(a)=[\mathrm T_{n-1}(a),\mathrm L]
\end{align}
with the boundary condition $\mathrm T_0(a) =\widehat{\mathrm H}(a)$. The $n$-th term in the sum in Eq.~\eqref{HhatH} is of order $\mathcal
O(a^{n+1})$ so that one can easily work out an approximation for $\mathrm H(a)$ with arbitrary precision, e.g.,
\begin{align}
\mathrm H(a) &=\widehat{\mathrm H}(a) +\mathrm T_1(a) \left(1 +\frac12 \mathrm T_1(a)\right)
\left( \bar\beta(a)+\frac12 \widehat{\mathrm H}(a) \right)
\notag\\
&+\frac12 \mathrm T_2(a)\left( \bar\beta(a)+\frac12\widehat{\mathrm H}(a)   \right)^2 + \mathcal O(a^4)\,.
\end{align}
It can be checked that this expression coincides with that obtained in ref.~\cite[Eq.~(3.9)]{Braun:2017cih}~\footnote{The notations adopted here
and in ref.~\cite{Braun:2017cih} differ slightly. To facilitate a comparison we note
 that the operators $\mathrm T_n$ defined here satisfy the equation $[S_+, \mathrm T_n] =n [\mathrm T_{n-1},z_1+z_2]$.}.

The evolution kernel $\widehat{\mathrm H}(a)$  can be realized as an integral operator. It acts on a function of two real variables as
follows
\begin{align}\label{Hintform}
\widehat{\mathrm H}(a)\, f(z_1,z_2)&=\! A f(z_1,z_2)\!+\!\!
\int_+ h(\tau) f(z_{12}^\alpha,z_{21}^\beta),
&
\end{align}
where $A$ is a constant, $z_{12}^\alpha \equiv z_1\bar\alpha +z_2 \alpha$, $\bar \alpha\equiv 1-\alpha$,
and
\begin{align}
\int_+ \equiv \int_0^1d\alpha \int_0^{\bar\alpha} d\beta.
\end{align}
$\tau=\alpha\beta/\bar\alpha\bar\beta$ is called conformal ratio. The weight function $h(\tau)$ in Eq.~\eqref{Hintform} only depends
 on this particular combination of the variables $\alpha, \beta$ 
as a consequence of invariance properties of $\widehat{\mathrm H}$, Eq.~\eqref{ScanH}.

It is easy to find that the operators $\mathrm T_n$ take the form
\begin{flalign}\label{Tintform}
\mathrm T_n(a)f(z_1,z_2)&=\!
\int_+ \ln^n(1\!-\!\alpha\!-\!\beta) h(\tau) f(z_{12}^\alpha,z_{21}^\beta)\,,
&
\end{flalign}
that again agrees with the results of ref.~\cite{Braun:2017cih}. Note, that
this expression does not depend on the choice of the branch of the logarithm
defining the function $\mathrm L=\ln z_{12}$ in Eq.~\eqref{T(H)}, see Appendix~\ref{appendix:A} for more discussion.

\section{Anomalous dimensions vs  kernels}\label{sect:gammas}
First of all let us establish a connection between the eigenvalues of the operators $\mathrm H$ and $\widehat{\mathrm H}$. Since both of
them are integral operators of the functional form in Eqs.~\eqref{Hintform}, \eqref{Tintform}, both operators are diagonalized by functions
of the form $\psi_N(z_1,z_2)=(z_1-z_2)^{N-1}$, where $N$ is an arbitrary complex number. One may worry that the continuation of the
function $\psi_N$ for negative $z_{12}$ is not unique and requires  special care. But it does not matter for our analysis. Indeed,
$z_{12}^\alpha-z_{21}^\beta= (1-\alpha-\beta)z_{12}$ with $\alpha+\beta<1$, therefore the operators do not mix the regions
$z_{12}\gtrless0$. For definiteness let us suppose that
\begin{align}\label{psiN}
\psi_N(z_1,z_2)=\theta(z_{12}) z_{12}^{N-1}.
\end{align}
Let $\gamma(N), \widehat\gamma(N)$ be eigenvalues (anomalous dimensions) of the operators $\mathrm H$, $\widehat{\mathrm H}$ corresponding
to the function $\psi_N$, 
respectively,
\begin{align}
\mathrm H(a)\psi_N & =\gamma(N)\psi_N, \\
 \widehat{\mathrm H}(a)\psi_N & =\widehat\gamma(N)\psi_N.
\end{align}
The anomalous dimensions $\gamma(N), \widehat\gamma(N)$  are analytic functions of $N$ in the right complex half-plane, $\mathrm{Re}(N) >0$.
For integer even (odd) $N$,  $\gamma(N)$ gives the anomalous dimensions of the local (axial)vector operators~\footnote{As usual one has to
consider the operators of certain parity, $\mathcal O_\pm(z_1,z_2)=\mathcal O(z_1,z_2) \mp \mathcal O(z_2,z_1)$, then the functions
$\gamma_\pm(N)$ give the anomalous dimensions of local operators, for even and odd $N$ respectively.}.

Now let us note that the operator $\mathrm T(\mathrm H)$ acts on  $\psi_N$ as follows
\begin{flalign}
\mathrm T(\mathrm H)\psi_N(z_1,z_2) & =\sum_{n=0}^\infty \frac{\mathrm L^n}{n!} \left(\bar\beta(a)+\frac12 \gamma(N)\right)^n\!\psi_N(z_1,z_2)
&
\notag\\
&=
z_{12}^{\bar\beta(a)+\frac12 \gamma(N)}\psi_N(z_1,z_2)
\notag\\
&=\psi_{N+\bar\beta+\frac12\gamma(N)}(z_1,z_2).
\end{flalign}
Thus, it follows from Eq.~\eqref{HhatH} that the anomalous dimensions $\gamma(N)$ and $\widehat\gamma(N)$
satisfy the relation (cf. also Eq.~\eqref{eq:no1})
\begin{align}
\gamma(N)=\widehat \gamma\left(N+\bar\beta(a)+\frac12\gamma(N)\right).
\end{align}
This relation appeared first in refs.~\cite{Dokshitzer:2005bf,Basso:2006nk} as an generalization of the Gribov-Lipatov reciprocity
relation~\cite{Gribov:1972ri,Gribov:1972rt}. It was shown that the asymptotic expansion of the function $\widehat\gamma(N)$ for large $N$
is invariant under  the reflection $N\to -N-1$, see e.g., refs.~\cite{Basso:2006nk,Dokshitzer:2006nm,Beccaria:2009vt,Alday:2015eya}.  This
property strongly restricts harmonics sums which can appear in the perturbative expansion of the anomalous dimension
$\widehat\gamma(N)$~\cite{Beccaria:2009vt}. Explicit expressions for $\widehat\gamma(N)$ are known at four loops in QCD~\cite{Moch:2017uml}
and at seven loops in the ${\cal N}=4$ SYM, see refs.~\cite{Kotikov:2008pv,Beccaria:2009vt,Beccaria:2009eq,Marboe:2014sya,Marboe:2016igj}.

\subsection{Kernels from anomalous dimensions}

For large $N$ the anomalous dimension $\widehat \gamma(N)$ grows as $\ln N$.
This term enters with a coefficient $2 \Gamma_\text{cusp}(a)$ where $\Gamma_\text{cusp}(a)$ is the so-called cusp
anomalous dimension~\cite{Polyakov:1980ca,Korchemsky:1987wg} whose complete form is known to the four-loop order in QCD~\cite{Henn:2019swt,vonManteuffel:2020vjv}
and in ${\cal N} = 4$ SYM~\cite{Henn:2019swt}. In the planar limit of  ${\cal N} =4$ SYM, the cusp anomalous dimension is known beyond the four-loop order (e.g., as a special case of results in~\cite{Marboe:2014sya,Marboe:2016igj}), and in fact,
to any loop order from ref.~\cite{Beisert:2006ez}.
Thus, we write $\widehat\gamma(N)$ in the following form
\begin{align}\label{gammaNstructures}
\widehat\gamma(N)=2\Gamma_\text{cusp}(a) S_1(N) + A(a) + \Delta\widehat\gamma(N)\,,
\end{align}
where $S_1(N)=\psi(N+1)-\psi(1)$ is the harmonic sum responsible for the $\ln N$ behavior at large $N$, and $A(a)$ is a constant term. The
remaining term, $\Delta\widehat\gamma(N)$, vanishes at least as $O(1/N(N+1))$ at large $N$.
The constant $A(a)$ is exactly the same which appears in Eq.~\eqref{Hintform}. The first term in Eq.~\eqref{gammaNstructures} comes from a
special $\mathrm{SL}(2,\mathbb R)$ invariant kernel
\begin{align}
\widehat{\mathcal H}f &=\int_0^1 \frac{d\alpha}{\alpha}\Big\{2 f(z_1,z_2)-\bar\alpha\big(f(z_{12}^\alpha,z_2)+ f(z_1,z_{21}^\alpha)\big)\Big\},
\end{align}
which in momentum space gives rise to the so-called plus-distribution.
The eigenvalues of this kernel are $2 S_1(N)$ ($\widehat{\mathcal H} z_{12}^{N-1}=2 S_1(N)z_{12}^{N-1}$).  It corresponds to a singular
contribution of the form $-\delta_+(\tau)$ to the invariant kernel $h(\tau)$, see ref.~\cite[Eq.~(2.19)]{Braun:2017cih} for detail. Thus
the evolution kernel can be generally written as
\begin{align}\label{Hhatform}
\widehat{\mathrm H} & =\Gamma_\text{cusp} (a)\widehat{ \mathcal H}+ A(a) +\Delta \widehat{\mathrm H}\,.
\end{align}
Here $\Delta \widehat{\mathrm H}$  is an  integral operator,
\begin{align}\label{DeltaH}
\Delta \widehat{\mathrm H}f(z_1,z_2)&=\int_+  h(\tau) f(z_{12}^\alpha,z_{21}^\beta)\,,
\end{align}
where the weight function $ h(\tau)$ is a regular function of $\tau\in (0,1)$. The eigenvalues of $\Delta \widehat{\mathrm H}$ are equal to
$\Delta\widehat\gamma(N)$ and are given by the following integral
\begin{align}\label{Deltagammah}
\Delta\widehat\gamma(N)
=\int_+  h(\tau) (1-\alpha-\beta)^{N-1}\,.
\end{align}
The inverse transformation takes the form~\cite{Braun:2014vba}
\begin{align}\label{inverseTransform}
 h(\tau)&=\int_C \frac{dN}{2\pi i} (2N+1) \Delta\widehat\gamma(N)P_N\left(\frac{1+\tau}{1-\tau}\right),
\end{align}
where $P_N$ are the Legendre polynomials. The integration path $C$ goes along the line parallel to the imaginary axis,  $\text{Re} (N)>0$,
such that all poles of $\Delta\widehat\gamma(N)$ lie to the left of this line. Some details of the derivation can be found in Appendix
\ref{appendix:B}.

One can hardly hope to evaluate the integral~\eqref{inverseTransform} in a closed form for an arbitrary function $\Delta\widehat\gamma(N)$.
However, as was mentioned before, the anomalous dimensions $\Delta
\widehat\gamma(N)$ in quantum field theory are rather special functions. Most of the
terms in the perturbative expansion of $\Delta\widehat\gamma(N)$  have  the following form
\begin{align}\label{etakomega}
\eta^k(N)\,\Omega_{\vec{m}}(N), &&  \eta^k(N)\,\Omega_1^p(N)
\end{align}
where $\eta(N)=1/(N(N+1))$, and the functions $\Omega_{\vec{m}}(N)=\Omega_{{m_1,\ldots,m_p}}(N)$ are the parity respecting harmonic
sums~\cite{Beccaria:2009vt}, ($\Omega_{\vec m}(N) \sim \Omega_{\vec m}(-N-1)$ for $N\to \infty$). We will assume that the sums
$\Omega_{\vec{m}}(N)$ are ``subtracted", i.e. $\Omega_{\vec{m}}(N)\to 0$ at $N\to\infty$. The second structure occurs only for $k>0$, since
$\Omega_1(N)=S_1(N)$ grows as $\ln N$ for large $N$.

Since all  $\mathrm{SL}(2,\mathbb R)$ invariant operators share the same eigenfunctions, the product of two invariant operators $H_1$ and
$H_2$, $H_1 H_2 ( =H_2H_1)$ with eigenvalues $H_1(N)$ and $H_2(N)$ respectively, has eigenvalues $H_1(N)H_2(N)$. One can use this property
to reconstruct an operator with the eigenvalue~\eqref{etakomega}.

First, we remark that the operator with the eigenvalues $\eta(N)$, (we denote it as $\mathcal H_+$), has (as follows from
Eq.~\eqref{Deltagammah}) a very simple weight function, $h_+(\tau)=1$. This can also be derived from Eq.~\eqref{inverseTransform}. Since
$P_N(x)=P_{-N-1}(x)$ the integral in Eq.~\eqref{inverseTransform} vanishes for the integration path $\mathrm{Re} (N)=-1/2$ due to
antisymmetry of the integrand. Therefore, the integral~\eqref{inverseTransform} can be evaluated by the residue theorem~\footnote{This
trick allows one to calculate the integral~\eqref{inverseTransform} for any function $\Delta\widehat\gamma(N)$ with {\it exact} symmetry
under $N\to -1-N$ reflection.}
\begin{align}
h_+(\tau)=\frac{2N+1}{N+1}P_N\left(\frac{1+\tau}{1-\tau}\right)\Big|_{N=0}=1.
\end{align}
\vskip2mm

Let us consider the product $H_2 = H_+\, H_1 (=H_1\, H_+)$, where $H_1$ is an integral operator with the weight function $h_1(\tau)$. Then
the weight function $h_2(\tau)$ of the operator $H_2$ is given by the following integral
\begin{align}\label{hplusconvolution}
h_2(\tau)= \int_0^\tau \frac{ds}{\bar s^2} \ln(\tau/s) h_1(s),
\end{align}
see Appendix~\ref{appendix:B} for details.  Thus the contribution to the anomalous dimension of type~\eqref{etakomega} can be evaluated
with the help of this formula if the weight function corresponding to the harmonic sums $\Omega_{\vec{m}}$ is known.

We also give an expression for another product of the operators: $H_2= \widehat{\mathcal H} H_1$,
\begin{align}\label{hhatconvolution}
h_2(\tau) &= -\ln \tau \, h_1(\tau) +2\bar\tau \int_0^{\tau} \frac{ds}{\bar s}\frac{h_1(\tau)-h_1(s)}{(\tau-s)}\,,
\end{align}
which appears to be useful in the calculations as well.

\subsection{Recurrence procedure}\label{subs:rec}

Let us consider the integral~\eqref{inverseTransform} with $\Delta\widehat\gamma =\Omega_{\vec{m}}$,
\begin{align}\label{hOmega}
h_{\vec{m}}(\tau)&=\int_C \frac{dN}{2\pi i} (2N+1)\Omega_{\vec{m}}(N)P_N\left(z\right),
\end{align}
where $z=(1+\tau)/(1-\tau)$. Using a recurrence relation for the Legendre functions
\begin{align}
(2N+1) P_N(z) & =\frac{d}{dz}\Big( P_{N+1}(z) - P_{N-1}(z)\Big)
\end{align}
we obtain
\begin{align}\label{hdzF}
h_{\vec{m}}(\tau)& =-
\frac d{dz} \int_C \frac{dN}{2\pi i} P_N(z) F_{\vec{m}}(N),
%
\end{align}
where
\begin{align}
F_{\vec{m}}(N)&=\Big( \Omega_{\vec{m}}(N+1)-\Omega_{\vec{m}}(N-1)\Big).
\end{align}
It is easy to see that the function $F_{\vec{m}}(N)$ has the negative parity under $N\to -N-1$ transformation and can be represented in the
form
\begin{align}\label{Fomega}
F_{m_1,\ldots, m_p}(N) &=\sum_{k=2}^p r_k(N) \Omega_{m_k,\ldots,m_p}(N) + r(N),
\end{align}
where $r_k(N)$ are rational functions of $N$. The harmonic sums $\Omega_{m_k,\ldots,m_p}(N)$ in Eq.~\eqref{Fomega} can be  either of
positive or negative parity. Therefore the coefficient $r_k(N)$ accompanying  the positive parity function $\Omega_{m_k,\ldots,m_p}(N)$ has
the form $r_k(N)=(2N+1) P_k(\eta)$, where $P_k$ is some polynomial, while $r_k=P_k(\eta)$ for the harmonic sums of negative parity. The
free term has the form $r(N)=(2N+1) P(\eta)$. Together, they make  $F_{m_1,\ldots, m_p}(N)$ with negative parity. For example, for the
harmonic sum $\Omega_{1,3}$ (see appendix \ref{appendix:kernels} for a definition), one gets
\begin{align}\label{F13}
F_{1,3}(N) &=(2N+1)\underline{\eta\left(\Omega_3+\zeta_3  -\eta^2-\frac12\eta^3\right)},
\end{align}
while for the harmonic sum $\Omega_{2,2}$
\begin{align}\label{F22}
F_{2,2}(N) &=(2N+1)\frac12\eta^3(3+\eta) +\underline{ \eta(2+\eta)\Omega_2 }.
\end{align}
Note the reappearance of the common factor $(2N+1)$ in the first case, \eqref{F13}. This implies that, up to the derivative $d/dz$, the
integral~\eqref{hdzF} has the form~\eqref{inverseTransform}.
Hence, if the kernel corresponding to the underlined terms in Eq.~\eqref{F13} is known, the kernel corresponding to $\Omega_{1,3}$ can be
easily obtained. Thus the problem of finding the invariant kernel with the eigenvalues $\Omega_{1,3}(N)$ is reduced to the problem of
finding the kernel with the eigenvalues $\Omega_3(N)$ $( \Omega_{1,3}\mapsto\Omega_{3} )$.

 However, as it seen from our second example, not all parity
preserving harmonic sums share this property. Indeed, the underlined term on
the right hand side (rhs) of Eq.~\eqref{F22} does not have the factor $(2N+1)$.
Hence, all these transformations do not help  to solve  the problem for $\Omega_{2,2}$.

It is easy to see that the above recurrence procedure works only if all the harmonic sums $\Omega_{m_k,\ldots,m_p}$ appearing in
Eq.~\eqref{Fomega} are  of positive parity. It was proven in ref.~\cite[Theorem 2]{Beccaria:2009vt} that  any harmonic sum,
$\Omega_{\vec{m}}$, with all indices $\vec m$ positive odd or negative even has positive parity (see Appendix~\ref{appendix:kernels} for
explicit examples of the harmonic sums satisfying these conditions). Therefore, the rhs of Eq.~\eqref{Fomega} only contains harmonic sums
of the same type.  Thus the invariant kernels corresponding to the harmonic sums of positive parity can \textit{always} be calculated
recursively, using Eqs.~\eqref{hdzF}, \eqref{Fomega} and~\eqref{hplusconvolution}, \eqref{hhatconvolution}. Crucially, only such harmonic
sums appear in  the anomalous dimensions~$\widehat \gamma(N)$ in QCD and ${\cal N}=4$ SYM. All convolution
integrals~\eqref{hplusconvolution} and \eqref{hhatconvolution} can in turn be systematically calculated with the packages
HyperInt~\cite{Panzer:2014caa} or PolyLogTools~\cite{Duhr:2019tlz}.

The explicit expressions for the kernels corresponding to the lowest harmonic sums are given in Appendix~\ref{appendix:kernels} for references.

\subsection{Invariant kernels: QCD}\label{section:qcd}

Below we give an explicit expression for the invariant kernel of twist-two flavor nonsinglet operator in QCD. We will not split the
operator $\mathcal O(z_1,z_2)$ into positive (negative) parity operators. The evolution operator still takes the form~\eqref{Hhatform},
with $\Delta\widehat{\mathrm H}$  given by the following integral
\begin{align}\label{DeltaHP}
\Delta \widehat{\mathrm H}f(z_1,z_2)&=\int_+ ( h(\tau) +  \bar h (\tau)P_{12} ) f(z_{12}^\alpha,z_{21}^\beta)\,,
\end{align}
where $P_{12}$ is a permutation operator, $ P_{12} f(z_1, z_2)= f(z_2,z_1) $\footnote{ In order to avoid possible misunderstandings we
write down it explicitly, $P_{12}  f(z_{12}^\alpha,z_{21}^\beta) =f(z_{21}^\alpha,z_{12}^\beta)$.}. For (anti)symmetric functions
$f(z_1,z_2)$ the operator~\eqref{DeltaHP} takes a simpler form \eqref{DeltaH} with the kernel $ h\pm \bar h$.

Our expression for the constant  term $A(a)$ agrees with the constant term $\chi$ given in ref.~\cite[Eq.~(5.5)]{Braun:2017cih}, $A =\chi
-2\Gamma_\text{cusp}$. For completeness, we provide explicit expressions for the constant $A = a A_1+a^2 A_2 + a^3 A_3 +\cdots$,
\allowdisplaybreaks{
\begin{widetext}
\begin{align}
A_1&= -6 C_F\, , \notag\\
A_2&= C_F\left[n_f\left(\frac{16}3\zeta_2+\frac23\right)-N_c\left(\frac{52}{3}\zeta_2+\frac{43}{6}\right)
+\frac{1}{N_c}\left(24\zeta_3-12\zeta_2+\frac32\right)\right]\, ,
\notag\\
A_3&=
C_F\bigg[n_f^2\left(\frac{32}9\zeta_3-\frac{160}{27}\zeta_2+\frac{34}9\right)
+n_fN_c\left(-\frac{256}{15}\zeta_2^2+\frac89\zeta_3+\frac{2492}{27}\zeta_2-17\right)
+ \frac{n_f}{N_c}\left(\frac{232}{15}\zeta_2^2-\frac{136}{3}\zeta_3+\frac{20}{3}\zeta_2-23\right)
\notag\\
&\qquad+ N_c^2\left(-80\zeta_5+\frac{616}{15}\zeta_2^2+\frac{266}9\zeta_3
-\frac{5545}{27}\zeta_2+\frac{847}{18}\right) +\left(-120\zeta_5-16\zeta_2\zeta_3
-\frac{124}{15}\zeta_2^2+\frac{1048}3\zeta_3-\frac{356}{3}\zeta_2+\frac{209}{4}\right)
\notag\\
&\qquad+\frac{1}{N_c^2}\left(120\zeta_5+16\zeta_2\zeta_3
-\frac{144}{5}\zeta_2^2-34\zeta_3-9\zeta_2-\frac{29}{4}\right)\bigg]\, ,
\end{align}
\end{widetext}
where $C_F=(N_c^2-1)/(2N_c)$ is the quadratic Casimir in the fundamental representation of $SU(N_c)$ and we take $T_F=1/2$.
Note that we are adopting a different color basis compared to ref.~\cite{Braun:2017cih}.
}

 The explicit expressions for the cusp anomalous dimensions $\Gamma_{\rm cusp}(a) = a \Gamma_{\rm cusp}^{(1)}  + a^2 \Gamma_{\rm
cusp}^{(2)} + a^3 \Gamma_{\rm cusp}^{(3)}$ up to three loops are provided in Eq.~\eqref{eq:cusps}.
Finally
we give answers for the kernels $h(\bar h)(a)=\sum_k a^k h_k (\bar h_k)$. Explicit one- and two-loop expressions are
known~\cite{Braun:2014vba,Braun:2017cih} but for completeness we give them here
\begin{align}
h_1=-4C_F\,, && \bar h_1=0\,,
\end{align}
and
\begin{align}
h_2&=C_F\biggl\{ n_f \frac{88}9 + N_c\left(-2 \mathrm H_1 + 8\zeta_2 -\frac{604}9 \right)
\notag\\
&\quad
+\frac1{N_c}\left( - 8\Big(\mathrm H_{11} +\mathrm H_2\Big)
  +  2 \left(1-\frac 4 \tau\right) \mathrm H_1\right)\biggr\}\,,
     \notag\\
\bar h_2 &=-\frac{8 C_F}{N_c}\left(\mathrm H_{11} + \tau\,\mathrm H_1\right)\,,
\end{align}
where $\mathrm H_{\vec{m}}=\mathrm H_{\vec{m}}(\tau)$ are the harmonic polylogarithms (HPLs)~\cite{Remiddi:1999ew}. The three-loop
expression\footnote{A file with our main results can be obtained from the preprint server
{\tt \href{http://arXiv.org} http://arXiv.org} by downloading the source. Furthermore, they are
available from the authors upon request.} is more involved {\allowdisplaybreaks
\begin{widetext}
\begin{align}
h_3&=
C_F\biggl\{ -\frac{64}9 n_f^2
+n_f N_c\frac83\biggl[
 \mathrm H_3 -\mathrm H_{110} -\mathrm H_{20} +\mathrm H_{12}
+\frac1\tau\mathrm H_{2} - \frac1\tau\mathrm H_{10}
-\frac{19}{12}\mathrm H_1 + 8 \zeta_3 -\frac{32}3\zeta_2
+\frac{5695}{72}\biggr]
\notag\\
&\quad
+\frac{n_f}{N_c}\frac{16}3\biggr[3\zeta_3 -\frac{75}{16}+\mathrm H_{3} + \mathrm H_{21}+ \mathrm H_{12} +\mathrm H_{111}
+\left(\frac{16}3+\frac1\tau\right) \Big( \mathrm H_2+\mathrm H_{11}\Big)
 +
 \left(\frac{31}{24}+\frac{10}{3\tau}\right)\mathrm H_1\biggr]
\notag\\
&\quad
+N_c^2 4\biggl[   \mathrm H_{13}
                  + \mathrm H_{112}
                  -  \mathrm H_{120}
                  - \mathrm  H_{1110}
                           + 2 \mathrm H_4
                           -2 \mathrm H_{30}
                           -2 \mathrm H_{210}
                           +2 \mathrm H_{22}
                           + \left(\frac83-\frac2\tau\right) \Big(\mathrm H_{20}-\mathrm H_3
                                            +\mathrm H_{110}
                                            -\mathrm H_{12}\Big)
\notag\\
                           &\quad
                           - \frac54\Big(\mathrm H_{10}
                                    +\mathrm H_{11}\Big)
                           +\frac 2{3\tau}\Big(\mathrm H_{10}
                                            -\mathrm H_2 \Big)
                                          -\frac52\mathrm H_0
                                            +\left(\frac{115}{72}+\zeta_2+\frac1{\tau}\right)\mathrm H_1
              -\frac{44}{5}\zeta_2^2
              -\frac{22}{3}\zeta_3
              +\frac{436}{9}\zeta_2
               -\frac{4783}{27}
\biggr]
\notag\\
&\quad
+16\,\biggl[\mathrm H_4 -\mathrm H_{30}+\mathrm H_{13} +\mathrm H_{121}
                          -\frac32 \mathrm H_{120}
                          +\frac32\mathrm H_{22}
                          +\frac32\mathrm H_{112}
                          +2\mathrm H_{31}
                          +2\mathrm H_{1111}
                          +3\mathrm H_{211}
                          -\frac12\mathrm H_{1110} - \left(\frac{1}\tau + 1\right)\mathrm H_{20}
\notag\\
&\quad
                          -\left(\frac{11}6 - \frac{1}\tau\right)\mathrm H_3
                          - \mathrm H_{110}
                          +\left(-\frac{37}{12} + \frac{3}{2\tau}\right) \mathrm H_{12}
                          -\left(\frac{7}{3} - \frac2\tau\right) \mathrm H_{21}
                          +\left( -\frac{43}{12} + \frac{3}\tau\right)\mathrm H_{111}
                          +\Big(\frac{13}8+ \frac12\zeta_2\Big)\mathrm H_{10}
\notag\\
&\quad                          -\left(\frac12\zeta_2 + \frac{127}9 + \frac{11}{6\tau}\right) \mathrm H_2
                          -\left(\frac{899}{72} + \frac{1}{3\tau}\right) \mathrm H_{11} + \Big(\zeta_2-1\Big) \mathrm H_0
                          +\left(\frac74\zeta_2-\frac{143}{36} -\frac1{\tau}\left(\frac12\zeta_2+\frac{67}9\right)\right)\mathrm H_1
                          +\frac52\zeta_2
                          -\frac{47}{24}\biggr]
                          \notag\\
&\quad
+\frac8{N_c^2}\biggl[
                            \mathrm H_{4}
                          -\mathrm H_{30}
                           -\mathrm H_{210}
                           +\mathrm H_{112}
                           -\mathrm H_{1111}
                           -2\,\mathrm H_{120}
                           +2\,\mathrm H_{13}
                           +2\mathrm H_{31}
                           -2\mathrm H_{1110}
                           -2\mathrm H_{211}
                           +3\mathrm H_{121}
\notag\\
&\quad -\left(\frac12+\frac1 \tau\right)\Big( \mathrm H_{20}-\mathrm H_{3}+\mathrm H_{110} \Big)
            +2\left(1 + \frac1\tau\right)\,\mathrm H_{21}
            +\left(\frac32 - \frac2 \tau\right)\,\mathrm H_{111}
    +\left(\frac78 + \frac{3}{2 \tau}\right)\, \mathrm H_{10}
        -\left(\zeta_2-\frac12+\frac3{2 \tau}\right)\, \mathrm H_{2}
\notag\\
&\quad
        +\Big(\frac{11}8 -  \zeta_2\Big)\, \mathrm H_{11}
  -\frac{11}4 \, \mathrm H_{0}
          +\left( \zeta_2 -\frac{107}{16} -\frac{\zeta_2}\tau -\frac1{ 2\tau}\right)\, \mathrm H_{1}
          + \frac72
\biggr]
 \biggr\}
 \\
\intertext{and}
\bar h_3 & =
-8C_F\Biggl\{ -\frac{2n_f}{3N_c}\left[
\mathrm H_{111} +\mathrm H_{110}+\tau\,\mathrm H_{10} +\left(\frac{16}3 +\tau\right)\mathrm H_{11}
+\left(\frac12+\frac{10}3 \tau\right)\mathrm H_1 +\frac12
\right]
\notag\\
&\quad
                        +\mathrm H_{120}
                          +  \mathrm H_{22}
                          -\mathrm H_{1110}
                          -\mathrm H_{112}
                          -2\mathrm H_{121}
                          +2\mathrm H_{211}
                          -4\mathrm H_{1111}
                            +  \tau \mathrm H_{20}
                           +\left(\frac{13}6- \tau\right) \mathrm H_{110}
                          +\Big(\frac12 - 2 \tau \Big) \mathrm H_{12}
\notag\\
&\quad
                          + \Big(\frac{5}2-2 \tau\Big)\mathrm H_{21}
                          +\left(\frac{43}{6}-  6 \tau\right) \mathrm H_{111}
                          -\left(\zeta_2-\frac{13}6 \tau\right)\mathrm H_{10}
                          -\Big(3+\zeta_2 + \frac32 \tau \Big)\mathrm H_2
                          +\left(\frac{236}9+\frac{2}3 \tau\right)\mathrm H_{11}
                         -\zeta_2 \tau \mathrm H_0
                         \notag\\ &\quad
                          +\left(\frac{53}6 + \zeta_2 + 3 \zeta_3  + \frac{134}9\tau  + \zeta_2 \tau \right) \mathrm H_1
                         +\frac{11}6 +3\left(\zeta_2+\zeta_3-\frac12\right) \tau
\notag\\
&\quad
+\frac1{N_c^2}\biggl[ \mathrm H_{1111}
                            - \mathrm H_{22}
                            - \mathrm H_{211}
                            +3 \mathrm H_{120}
                            +  3    \mathrm H_{112}
                             -3  \mathrm H_{1110}
                                  +4  \mathrm H_{121}
                     +3\,\tau\,\mathrm H_{20}
                              +3\,\left(\frac12-\tau\right)\,\mathrm H_{110}
                              -\left(\frac72 -  4 \tau\right)\,\mathrm H_{12}
\notag\\
& \quad
                              + \left(\frac12 + 4 \tau\right)\, \mathrm H_{21}
                              +\left(-\frac32 + 2\tau\right)\,\mathrm H_{111}
                              -3\,\left(\zeta_2 - \frac12\tau\right)\, \mathrm H_{10}
                              +\left(\zeta_2 -2- \frac32 \tau\right)\, \mathrm H_{2}
                              +2\Big(\zeta_2  -1\Big)\, \mathrm H_{11}
                               -3\zeta_2\,\tau\,\mathrm H_{0}
\notag\\
&\quad
                              +\Big(5 +2\zeta_2 +3\zeta_3+  \zeta_2 \tau\Big)\,\mathrm H_{1}
                              +3\tau\left( \zeta_3- \frac12 \right)
\biggr]
\Biggr\}.
\end{align}
\end{widetext}
}
The kernels are smooth functions of $\tau$ except for the endpoints $\tau=0$ and $\tau=1$. For $\tau\to 1$ the three-loop kernel functions
behave as $ \sum_{0\leq k\leq 4}\sum_{m>0} r_{km}\bar\tau^m \ln^k\bar\tau $. For small $\tau$ -- which determines the large $N$ asymptotic
of the anomalous dimensions -- the kernels (for each color structure) have the form $\sum_{k\geq 0} (a_k + b_k\ln \tau) \tau^k $. We note
here that the reciprocity property of the anomalous dimension is equivalent to the statement that the small $\tau$ expansion  of the
kernels does not involve non-integer powers  of $\tau$, namely $ h(\tau)\sim\sum_{m,k\geq0} a_{mk}\tau^m \ln^k\tau $.

 Below we compare our exact three-loop results with the approximate
 expressions constructed in ref.~\cite{Braun:2017cih}. The approximate
 expressions reproduce the asymptotic behaviors of the exact kernels at both
 $\tau\to 0,1$. We therefore subtract the logarithmically divergent pieces
 (see Eqs.~\eqref{eq:asy0} and \eqref{eq:asy1} for explicit expressions) from
 both the exact and the approximated expressions to highlight their (small)
 deviations as shown in Figs.~\ref{diag:h3} and~\ref{diag:h3b}. For
 illustrative purposes, we plot the planar contribution ($C_FN_c^2$ and $C_F$
 in $h_3$ and $\bar h_3$ respectively) and the subsubplanar contribution
 ($C_F/N_c^2$). The former is numerically dominant and generates the leading
 contribution in the large-$N_c$ limit whereas the latter shows the worst-case
 scenario for the previous approximation using a simple HPL function
 ansatz. The error of other color structures all fall between the planar and
 subsubplanar cases, hence are numerically small.

%
\onecolumngrid
\begin{widetext}
\begin{figure}[!h]
\begin{center}
\includegraphics[width=0.9\textwidth]{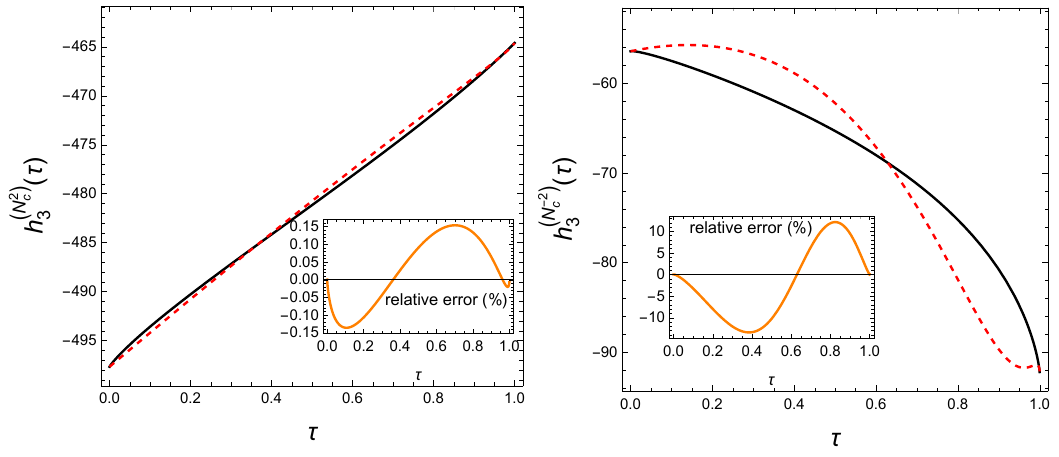}
\end{center}
\caption{Comparison of two distinct color contributions ($C_FN_c^2$ and
  $C_F/N_c^2$) in the exact (black solid) and approximated (red dashed)
  three-loop kernel $h_3$ (see ref.~\cite{Braun:2017cih} for explicit expressions of the latter).
   The inset curves show the relative percentage errors ($(1-h_{3,{\rm appr}}^{(c)}/h_{3,{\rm exact}}^{(c)})\times 100\%$) of the approximation.}
\label{diag:h3}
\end{figure}
%
%
\onecolumngrid
\begin{figure}[h]
\begin{center}
\includegraphics[width=0.9\textwidth]{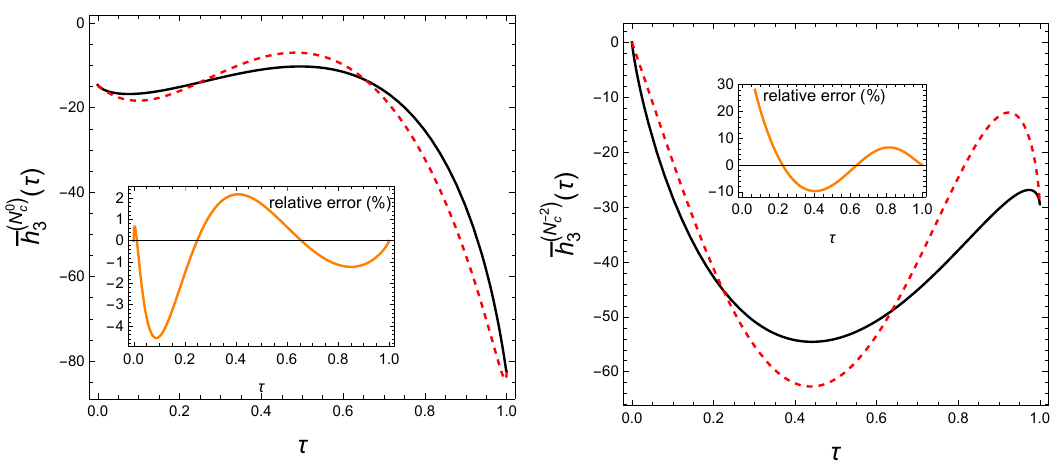}
\end{center}
\caption{Same as Fig.~\ref{diag:h3} for the color structures $C_F$ and $C_F/N_c^2$ in $\bar h_3$.
}
\label{diag:h3b}
\end{figure}
\end{widetext}
\twocolumngrid

\subsection{Invariant kernels: ${\cal N}=4$ SYM }\label{sect:SYM}
 In this section we present the invariant kernels for the universal anomalous
 dimensions of the planar ${\cal N}=4$ SYM, see e.g., refs.~\cite{Kotikov:2003fb,Kotikov:2008pv}
 for expressions up to NNLO.
 They are rather short so that we quote them here.
 We use the parametrization~\eqref{gammaNstructures},
 where $\Gamma_\text{cusp}(a)$ can be found in ref.~\cite{Henn:2019swt}
 and the constant term $A(a)$ is
\begin{align}
A(a)=-24 a^2 \zeta_3 + 32 a^3 \big(\zeta_2\zeta_3 +5\zeta_5\big) +O(a^4),
\end{align}
where $a= \frac{N_c g^2_\text{SYM}}{16\pi^2}$, and
\begin{align}
\Delta\widehat\gamma(N) &= - a^2 16\Big(\Omega_3 - 2 \, \Omega_{-2,1}  + 2\, \Omega_1 \, \Omega_{-2}\Big)
\notag\\
&\quad
+a^3  {64}\Big(
   \Omega_5
          +2\,\Omega_{3,-2}
          -8\,\Omega_{1,1,-2,1}+2\Omega_{1,-4}
\notag\\
&\quad
          + \Omega_1(\Omega_{-4}
          +\Omega_{-2}^2 
          +\zeta_2\Omega_{-2} ) \! -2\zeta_2 \Omega_{-2,1}
 \Big).
\end{align}
For the kernels we find $h_1=\bar h_1=0$,
\begin{align}
h_2 &=8 \frac{\bar \tau} \tau \mathrm H_{1}\,, & \bar h_2 &=- 8 {\bar \tau}  \mathrm H_{1}
\end{align}
and
\begin{align}
h_3 &= - 16\frac{\bar \tau} \tau \Big( 4\,\mathrm H_{1 1 1}  + \mathrm H_{21} + \mathrm H_{1 2} + \mathrm H_{1 1 0}\Big)\,,
\\
\bar h_3&= 16 \bar \tau \Big( 4\mathrm H_{1 1 1} + 3\big( \mathrm H_{21} +  \mathrm H_{1 2} \big)
- \mathrm H_{1 1 0} + \mathrm H_{2 0} - \zeta_2 \mathrm H_0\Big).\notag
\end{align}
These expressions are extremely simple in comparison with the  expressions in
QCD of the same order. Let us notice that the two-loop kernels contain only
HPLs of weight one with the three-loop kernels involving HPLs of weight three,
while in QCD the corresponding kernels require HPLs of weight two and four,
respectively. Note also that the kernel $h$ is proportional to the factor
$\bar \tau/\tau$ and the kernel  $\bar h$ to the factor $\bar \tau$.
It would be interesting to see if these properties persist in higher loops.

\section{Summary}\label{sect:summary}
We have constructed a transformation that brings the evolution kernels of twist-two operators to the canonically conformal invariant form. The
eigenvalues of these kernels are given by the parity respecting anomalous
dimensions. We have developed a recurrence procedure that allows one
to restore the weight functions of the corresponding  kernels. It is applicable to a subset of the harmonic sums (with positive odd and negative
even indices). It is interesting to note that exactly only such harmonic sums appear in the expressions for the reciprocity respecting anomalous
dimensions.

We have calculated the three-loop invariant kernels in QCD and in ${\cal N}=4$ SYM  (in the planar limit). In QCD it was the last missing
piece to obtain the three-loop evolution kernels for the flavor-nonsinglet twist-two operators in a fully analytic form, see
ref.~\cite{Braun:2017cih}.

In the case of ${\cal N}=4$ SYM the lowest order expressions for the kernels are rather simple and exhibit some regularities, $h\sim
\bar\tau/\tau$, $\bar h\sim\bar \tau$. It would be interesting to check if these properties survive at higher loops. We expect that at
$\ell$-loops the kernels $h^{(\ell)}(\tau)$ will be given by linear combinations (up to common prefactors) of HPLs of weight $2\ell -3$
with positive indices. Therefore going over to the invariant kernel can lead to a more compact representation of the anomalous dimensions
than representing the anomalous dimension spectrum $\gamma(N)$ in terms of harmonic sums. The much smaller function basis in terms of HPLs
($\tau/\bar\tau H_{\vec m}$ and $\bar\tau H_{\vec m}$) opens the possibility of extracting the analytical expressions of the higher-order
evolution kernels from minimal numerical input through the PSLQ algorithm.

\section*{Acknowledgments}
We are grateful to Vladimir M. Braun and Gregory P. Korchemsky for illuminating discussions and comments on the manuscript.
This work is supported by Deutsche Forschungsgemeinschaft (DFG) through
the Collaborative Research Center TRR110/2, grant 409651613 (Y.J.),
and the Research Unit FOR 2926, project number 40824754 (S.M.).

\appendix
\renewcommand{\theequation}{\Alph{section}.\arabic{equation}}

\section*{Appendices}

\section{}\label{appendix:A}
In this appendix, we describe in detail the derivations  of some of the equations presented in section~\ref{sect:prelim}. Let us start with
Eq.~\eqref{TintS}. For the generator $\mathrm S_-(a)=S_-$ the statement is trivial. Next, making use of Eq.~\eqref{T(H)} for the operator
$\mathrm T(\mathrm H)$ and, taking into account that $\mathrm H(a)$ commutes
with the generators $\mathrm S_\alpha(a)$, one can write the
left hand side (lhs) of Eq.~\eqref{TintS} in the form
\begin{align}\label{STL}
\sum_{n=0}^\infty \frac1{n!}\mathrm L^n \mathrm S_\alpha(a) \mathrm X^n
\, ,
%
\end{align}
where $\mathrm X=\bar\beta(a)+\frac12 \mathrm H(a) $. Using the representation~\eqref{Smathrm} for the generators
and taking into account that $[S_0,\mathrm L]=1$ and $[S_+,\mathrm L]=z_1+z_2$ (we recall that $\mathrm L=\ln z_{12}$) one obtains
\begin{align}\label{SS}
\mathrm L^n \mathrm S_0 &= S_0 \mathrm L^n - n \mathrm L^{n-1} + \mathrm L^n X,
\notag\\
\mathrm L^n \mathrm S_+ &= S_+ \mathrm L^n + (z_1+z_2)\left(-n \mathrm L^{n-1} + \mathrm L^n  X\right).
\end{align}
Substituting these expressions back into Eq.~\eqref{STL} one finds that the contributions of the last two terms on the rhs of
Eq.~\eqref{SS} cancel each other. Hence Eq.~\eqref{STL} takes the form
\begin{align}
S_\alpha\sum_{n=0}^\infty \frac1{n!}\mathrm L^n  \mathrm X^n = S_\alpha \mathrm T(\mathrm H),
\end{align}
that finally results in Eq.~\eqref{TintS}. \vskip 3mm

Let us now show that the inverse to $\mathrm T(\mathrm H)$ has the form~\eqref{inverseT}.  The product $\mathcal I=\mathrm T^{-1}(\mathrm
H)\mathrm T(\mathrm H)$ can be written as
\begin{align}
%
\mathcal I
 &=
\sum_{n=0}^\infty \frac{(-1)^n}{n!} \mathrm L^n\left(\bar\beta(a)+\frac12\widehat{  \mathrm H}(a) \right)^n \mathrm T(\mathrm H).
\end{align}
Moving $\mathrm T(\mathrm H)$ to the left with help of the
relation~\eqref{TintH} and then using Eq.~\eqref{T(H)} for $\mathrm T(\mathrm H)$
one gets ($\mathrm X=\bar\beta(a)+\frac12 \mathrm H(a) $)
\begin{align*}
\mathcal I=\sum_{n=0}^\infty \frac{(-1)^n}{n!} \mathrm L^n \mathrm T(\mathrm H) \mathrm X^n
=\sum_{n,k=0}^\infty \frac{(-1)^n}{n!k!} \mathrm L^{n+k}  \mathrm X^{n+k} =1\,.
\end{align*}

Finally, we consider the product of  operators $\mathrm T$ with a differently defined function $\mathrm L$. Namely, let us take $\mathrm
T_\pm(\mathrm H) \equiv \mathrm T(\mathrm L_\pm,\mathrm H)$, where $L_\pm =\ln(z_{12}\pm i0)$ so that $\mathrm L_+-\mathrm
L_-=2\pi\theta(z_2-z_1)$. In order to calculate the product $\mathrm U=\mathrm T_+(H) \mathrm T_-(H)$ one proceeds as before: use
expansion~\eqref{T(H)} for $\mathrm T_+(H)$, move $\mathrm T_-(H)$ to the left
and then expand it into a power series. It yields
\begin{align}\label{mathrmU}
\mathrm U &=\sum_{n,k=0}^\infty \frac{(-1)^n}{n!k!} \mathrm L_+^{n}\mathrm L_-^k  \widehat{\mathrm X}^{n+k}
\, ,
\end{align}
where $\widehat{\mathrm X}=\bar\beta(a)+\frac12 \widehat{\mathrm H}(a) $.
Let $L_+=L_-+2\pi i\theta(z_2-z_1)$ one can get for the sum in Eq.~\eqref{mathrmU}
\begin{align*}
\mathrm U &=\sum_{m=0}^\infty \frac{(2\pi i \theta)^m}{m!}\widehat{\mathrm X}^m(a)
=1 - \theta
\left(1- e^{2\pi i \left( \bar\beta+\frac12\widehat{\mathrm  H}\right)}\right),
\end{align*}
where $\theta\equiv\theta(z_2-z_1)$. Since $S_{0,+}\theta(z_{21})\sim z_{21}\delta(z_{21})=0$ one concludes that $\mathrm U$ commutes with
the canonical generators $S_\alpha$ and hence $\mathrm U\widehat{\mathrm H}=\widehat{\mathrm H}\mathrm U$.

\section{}\label{appendix:B}
Let us check that the kernel $h(\tau)$ given by Eq.~\eqref{inverseTransform} has the eigenvalues $\Delta\widehat\gamma(N)$.
First, after some algebra, the integral in Eq.~\eqref{Deltagammah} can be brought to the following form
\begin{align}\label{gammaQ}
\Delta\widehat\gamma(N)=\int_1^\infty dt\,  h\left(\frac{t-1}{t+1}\right) \, Q_N(t)\,,
\end{align}
where $Q_N(t)$ is the Legendre function of the second kind~\cite{MR1243179}. Inserting $h$ in the form of Eq.~\eqref{inverseTransform} into
Eq.~\eqref{gammaQ} one gets
\begin{align}
\int_C\frac{dN'}{2\pi i}(2N'+1)\Delta\gamma(N')\int_1^\infty dt\, P_{N'}(t) \, Q_N(t)\, .
\end{align}
The $t$-integral of the product of the two Legendre functions gives~\cite{MR1243179}
\begin{align}
\left((N-N')(N+N'+1)\right)^{-1}.
\end{align}
Then closing the integration contour in the right half-plane one evaluates the
$N'$ integral with the residue theorem at $N'=N$ yielding the desired lhs of Eq.~\eqref{gammaQ}.

Finally, in order to verify Eq.~\eqref{hplusconvolution} one can check that the integral~\eqref{gammaQ} with the kernel $h_2$,
$\Delta\widehat\gamma_2(N)$, is equal to $\Delta\widehat\gamma_1(N)/N/(N+1)$. The simplest way to do it is to  substitute the Legendre
function in the form
\begin{align}
Q_N(t)=-\partial_t(1-t^2)\partial_t Q_N(t)/N/(N+1),
\end{align}
and perform integration by parts.

\section{}\label{appendix:kernels}
In this appendix, we collect the harmonic sums and the corresponding kernels which we have used.
We split them into two parts: the first one includes the harmonic sums $\Omega_{m_1,\ldots,m_k}$ such that  $\prod_i^k \mathrm{sign}(m_i)=1$.
\allowdisplaybreaks{
\begin{align}
  &\Omega_{3} = S_3-\zeta_3,
\notag\\
 & \Omega_{3,1} =S_{3,1}-\frac12 S_4 
-\frac3{10}\zeta_2^2
%
\notag\\
  &\Omega_{-2,-2} = S_{-2,-2} -\frac12 S_{4}+\frac12\zeta_2S_{-2}+\frac18\zeta_2^2,
\notag\\
  &\Omega_{1,3,1} =  S_{1,3,1}-\frac12 S_{1,4}-\frac12 S_{4,1} + \frac14 S_5 -\frac3{10}\zeta_2^2 S_1  + \frac34\zeta_5 ,
\notag\\
  &\Omega_{-2,-2,1} =  S_{-2,-2,1}-\frac12 S_{4,1}-\frac12 S_{-2,-3} + \frac14 \zeta_3 S_{-2} +\frac{5}{16}\zeta_5 ,
\notag\\[2mm]
&\Omega_{5} = S_5-\zeta_5.
\end{align}
Here $S_{\vec m}$ are the harmonic sums with argument $N$.
We define the sums of negative signature, $\prod_i^k \mathrm{sign}(m_i)=-1$,
with an additional sign factor:
\begin{align}
  &\Omega_{-2} =(-1)^N \left[S_{-2}+\frac{\zeta_2}2\right],
\notag\\
  &\Omega_{-2,1} =(-1)^N \left[ S_{-2,1} - \frac12 S_{-3} + \frac14\zeta_3\right],
\notag\\
&\Omega_{1,-2,1} = (-1)^N\left[S_{1,-2,1} - \frac12S_{1,-3} - \frac12 S_{-3,1}  + \frac14 S_{-4} \right.\notag\\
    &\quad\left. + \frac14 \zeta_3 S_1-\frac1{80}\zeta_2^2 \right],
\notag\\
  &\Omega_{-4,1} =(-1)^N\left[S_{-4,1}-\frac12 S_{-5} + \frac{11}8\zeta_5 -\frac12\zeta_2\zeta_3\right],
\notag\\
  &\Omega_{3,-2} = (-1)^N\left[ S_{3,-2}-\frac12 S_{-5} +\frac12\zeta_2 S_3 + \frac98 \zeta_5-\frac34\zeta_2\zeta_3\right],
\notag\\
  &\Omega_{1,1,-2,1} = (-1)^N \biggl[S_{1,1,-2,1} -\frac12 S_{1,1,-3}-\frac12 S_{1,-3,1}
\notag\\
   &\qquad\quad-\frac12 S_{2,-2,1} +\frac14 S_{2,-3} +\frac14 S_{-4,1}  +\frac14 S_{1,-4} -\frac18 S_{-5}
\notag\\
   &\qquad\quad +\frac14 \zeta_3 S_{1,1}-\frac1{80}\zeta_2^2 S_1 -\frac18 \zeta_3 S_2 + \frac{1}{8}\zeta_5 -\frac1{16}\zeta_2\zeta_3
   \biggr],
\notag\\
   &\Omega_{1,-4}=(-1)^N\biggl[S_{1,-4}-\frac12S_{-5}+\frac7{20}\zeta_2^2S_1-\frac{11}{8}\zeta_5+\frac12\zeta_2\zeta_3\biggr].
 \end{align}
These combinations of harmonic sums are generated by the following kernels,
\begin{align}
&{\mathcal H}_3
= -\frac12 \frac{\bar \tau} \tau \mathrm H_1,
\notag\\
& {\mathcal H}_{3,1} = \frac14 \frac{\bar \tau}{\tau}\left(\mathrm H_{11}+\mathrm H_{10}\right)
\notag\\
& {\mathcal H}_{-2,-2} = \frac14\frac{\bar\tau}{\tau}\mathrm H_{11},
\notag\\
&{\mathcal H}_{1,3,1} = -\frac18 \frac{\bar \tau} \tau \left(\mathrm H_{20}+\mathrm H_{110}+\mathrm H_{21} +\mathrm H_{111} \right),
\notag\\
& {\mathcal H}_{-2,-2,1} = \frac18 \frac{\bar \tau}\tau \left(\mathrm H_{12}-\mathrm H_{110}\right),
\notag\\
&\mathcal H_5
= -\frac12 \frac{\bar \tau} \tau \left(\mathrm H_{111}+\mathrm H_{12}\right)
\end{align}
and 
\begin{align}
&{\mathcal H}_{-2} = \frac12 \bar \tau,
\notag\\
&{\mathcal H}_{-2,1}   = -\frac{1}4 \bar\tau(\mathrm H_1+\mathrm H_0),
\notag\\
&{\mathcal H}_{1,-2,1} = \frac18 \bar \tau \left( \mathrm H_{10} + \mathrm H_{11}\right),
\notag\\
& {\mathcal H}_{-4,1} = -\frac14 \bar \tau \left(\mathrm H_{21}+\mathrm H_{20}+\mathrm H_{111}+\mathrm H_{110}\right),
\notag\\
&{\mathcal H}_{3,-2}  = -\frac14\bar \tau \left(\mathrm H_{21} +\mathrm H_{111}\right),
\notag\\
&{\mathcal H}_{1,1,-2,1} =-\frac1{16}\bar \tau \, \left(\mathrm H_{111}+\mathrm H_{110}\right),
\notag\\
&{\mathcal H}_{1,-4} =-\frac14\bar\tau\left(\mathrm H_{12}+\mathrm H_{111}\right),
\end{align}
where all HPLs have argument $\tau$.
These functions serve as a basis and more complicated structures can be generated as products of $\Omega_{\vec m}$.

}

\section{}\label{appendix:expansions}
Here we give the small ($\tau\to 0$) and large ($\tau\to 1)$ expansions of the invariant kernels $h_3,\bar h_3$. By $h_3^{(A)}$ ($\bar
h_3^{(A)}$) we denote the function which appears in the expression for $h_3$ ($\bar h_3^{(A)}$) with the color factor $C_F\times A$. We
will keep the logarithmically enhanced and constant terms in both limits. The former is subtracted from both the exact and approximated
three-loop kernel to obtain the two figures in Eqs.~\ref{diag:h3} and~\ref{diag:h3b}. At $\tau \to 0$ one gets \allowdisplaybreaks{
\begin{align}\label{eq:asy0}
h_3^{(n_f N_c)} & = \frac{5839}{27} - \frac{256}{9}\zeta_2 + \frac{64}3\zeta_3 -\frac83 \ln \tau\,,
\notag\\
h_3^{(n_f/N_c)} &= -\frac {17}9 + 16\zeta_3\,,
\notag\\
 \bar h_3^{(n_f/N_c)} &= \frac 8 3\,,
 \notag\\
h_3^{(N_c^2)} &= -\frac{18520}{27} - \frac{88}{3}\zeta_3 - \frac{176}{5}\zeta_2^2 + \frac{1744}{9}\zeta_2
                  -\frac{46}{3}\ln \tau
\notag\\
   h_3^{(N_c^0)} &=              -\frac{ 1186}9 + 32\zeta_2 + (-32+16\zeta_2)\ln\tau\,,
\notag\\
h_3^{(N_c^{-2})} &= 24-8\zeta_2 -18 \ln\tau\,,
\notag\\
 \bar  h_3^{(N_c^0)} &=  -\frac{44}3\,,
\notag\\
\bar h_3^{(N_c^{-2})} &= -48\tau\left(\zeta_2 + \zeta_3+\frac14  - \zeta_2 \ln\tau \right)\, ,
\end{align}
and for $\tau\to 1$ one obtains
\begin{align}\label{eq:asy1}
h_3^{(n_f N_c)} & = \frac{5695}{27}  - \frac{208}{9} \zeta_2 + \frac{64}{3} \zeta_3 + \left(-\frac{16}3\zeta_2+\frac{38}9\right) \ln\bar\tau\,,
\notag\\
h_3^{(n_f/N_c)} &= \frac{304}9\zeta_2 + 16\zeta_3-25 -\left(\frac{16}3\zeta_2 +\frac{74}3\right) \ln\bar\tau
\notag\\
&\quad + \frac{152}9\ln^2\bar\tau -\frac89\ln^3\bar\tau\,,
\notag\\
 \bar h_3^{(n_f/N_c)} &= \frac{16}3\left(\frac12-\zeta_2 + \zeta_3\right) + \left(\frac{16}3\zeta_2-\frac{184}9\right)\ln\bar\tau
\notag\\
&\quad
                        +\frac{152}9\ln^2\bar\tau -\frac89\ln^3\bar\tau\,,
\notag\\
h_3^{(N_c^2)} &=-\frac{72}5\zeta_2^2 + \frac {1741}9 \zeta_2 - \frac{88}3 \zeta_3 - \frac{19132}{27}
\notag\\
&\quad
		+\left(\frac 4 3 \zeta_2 -\frac{187}{18}\right) \ln\bar\tau
		+\left( -\frac{5}2 + 4\zeta_2\right)\ln^2\bar\tau\,,
		\notag\\
h_3^{(N_c^0)} &=    \frac{136}5\zeta_2^2 - \frac{2170}9\zeta_2 + 80\zeta_3 - \frac{94}3
\notag\\
		&\quad
		+\left( -\frac{32}3\zeta_2- 24\zeta_3 + \frac{548}3\right)\ln\bar\tau
		\notag\\
		&\quad
		+\left( 16\zeta_2-\frac{923}9\right) \ln^2\bar\tau
		+\frac{14}9 \ln^3\bar\tau
		+\frac 43 \ln^4\bar\tau\,,
\notag\\
		h_3^{(N_c^{-2})} &=-28\zeta_2^2 - 27\zeta_2 + 56\zeta_3 + 28
		\notag\\
		&\quad
		+\left(\frac{115}2 -12\zeta_2-40\zeta_3\right)\ln\bar\tau
		\notag\\
		&\quad
		+\left(8\zeta_2+\frac{11}2\right)\ln^2\bar\tau +\frac 23\ln^3\bar\tau -\frac 13\ln^4\bar\tau\,,
\notag\\
 \bar  h_3^{(N_c^0)} &= -\frac{136}5\zeta_2^2 + \frac{88}3\zeta_2 - \frac{136}3\zeta_3-\frac 83
 	            \notag\\
		    &\quad
		    +\left( -\frac{16}3\zeta_2 +\frac{1708}9\right) \ln\bar\tau
		     \notag\\
		    &\quad
		    -\frac{968}9\ln^2\bar\tau  +\frac{14}9 \ln^3\bar\tau +\frac 43 \ln^4\bar\tau\,,
\notag\\
\bar h_3^{(N_c^{-2})} &= -\frac{216}5\zeta_2^2 + 40\zeta_2 + 8\zeta_3 + 12
			 \notag\\
		          &\quad
		      +   (40\zeta_2 - 64\zeta_3+40)\ln\bar\tau
                          \notag\\
		          &\quad
		          -8(4\zeta_2-1)\ln^2\bar\tau  + \frac 23 \ln^3\bar\tau - \frac 13\ln^4\bar\tau\,.
\end{align}

}
Here we quote the cusp anomalous dimensions up to three loops for reference~\cite{Polyakov:1980ca,Korchemsky:1987wg,Moch:2004pa},
\begin{align}\label{eq:cusps}
\Gamma_{\rm cusp}^{(1)}&= 4C_F\, ,
\notag\\
\Gamma_{\rm cusp}^{(2)}&= C_F\left[N_c\left(\frac{268}{9}-8\zeta_2\right)-\frac{40}{9}n_f  \right]\, ,
\notag\\
\Gamma_{\rm cusp}^{(3)}&=
C_F\bigg[N_c^2\left(\frac{176}{5}\zeta_2^2+\frac{88}3\zeta_3-\frac{1072}9\zeta_2+\frac{490}{3}\right)
\notag\\
&\quad + N_c n_f\left(-\frac{64}{3}\zeta_3+\frac{160}{9}\zeta_2-\frac{1331}{27}\right)
\notag\\
&\quad
+\frac{n_f}{N_c}\left(-16\zeta_3+\frac{55}{3}\right)
-\frac{16}{27}n_f^2 \bigg]\, .
\end{align}


\bibliography{ymmbib}

\end{document}